\documentclass[10pt,a4paper,reqno]{amsart}
\usepackage{amsthm}
\usepackage{amsmath}
\usepackage{amsmath, amssymb}
\usepackage{type1cm}
\usepackage{cite}
\theoremstyle{definition}
\newtheorem{defi}{Definition}
\newtheorem{prop}{Proposition}
\newtheorem{lemma}{Lemma}
\newtheorem{theorem}{Theorem}
\newtheorem{corollary}{Corollary}
\newcommand{\eqdef}{\stackrel{\mathrm{def}}{=}}
\begin{document} 
\title{Divergence functions in dually flat spaces and their properties}
\author{Tomohiro Nishiyama}
\begin{abstract}
In the field of statistics, many kind of divergence functions have been studied as an amount which measures the discrepancy between two probability distributions. In the differential geometrical approach in statistics (information geometry), dually flat spaces play a key role. In a dually flat space, there exist dual affine coordinate systems and strictly convex functions called potential and a canonical divergence is naturally introduced as a function of the affine coordinates and potentials. The canonical divergence satisfies a relational expression called triangular relation. This can be regarded as a generalization of the law of cosines in Euclidean space.

In this paper, we newly introduce two kinds of divergences. 
The first divergence is a function of affine coordinates and it is consistent with the Jeffreys divergence for exponential or mixture families.  For this divergence, we show that more relational equations and theorems similar to Euclidean space hold in addition to the law of cosines. 
The second divergences are functions of potentials and they are consistent with the Bhattacharyya distance for exponential families and are consistent with the Jensen-Shannon divergence for mixture families respectively. We derive an inequality between the the first and the second divergences and show that the inequality is a generalization of Lin's inequality.\\

\smallskip
\noindent \textbf{Keywords:} information geometry, parallelogram law, polarization identity, Jeffreys divergence, Jensen-Shannon divergence, Renyi divergence,  Bhattacharyya distance.
\end{abstract}
\date{}
\maketitle
\bibliographystyle{plain}
\section{Introduction}
Given a probability distribution $p_\theta(x)$, we consider a manifold $M$ with a parameter vector $\boldsymbol{\theta}=(\theta^1,\theta^2,\cdots,\theta^n)\in \mathbb{R}^n$ as a coordinate system.
Manifolds with probability distributions as elements are called statistical manifolds, and dually flat space is an important concept of it \cite{amari2016information}.
In a dually flat space, it is possible to introduce a dual coordinate system $\boldsymbol{\eta}=(\eta_1,\eta_2,\cdots,\eta_n)$ and the convex functions called potential $\psi(\theta)$ and $\phi(\eta)$.

For example, the manifold of the exponential family $p_\theta(x)=\exp(C(x)+\sum_{i=1}^n\theta^iF_i(x)-\psi(\theta))$ is a dually flat space for $\boldsymbol{\theta}$ and $\boldsymbol{\eta}=E[\textbf{F}(x)]$, where $x$ is a random variable, $F_i(x), C(x)$ are known functions and $E[\cdot]$ denotes expected value.
Similarly, the manifold of the mixture family $p_\eta (x)=p_0(x)+\sum_i \eta_i(p_i(x) - p_0(x))$ is a dually flat space, where $p_i(x)(i=0,1,2,\cdots, n)$ are probability distributions.

Given two probability distributions, the Kullback-Leibler divergence(KL-divergence) has long been known as an amount which measures the discrepancy.
In a dually flat space, we may define an amount called a canonical divergence and the canonical divergence is consistent with the KL-divergence for exponential or mixture families \cite{amari2016information,amari2010information,amari2007methods,bregman1967relaxation}. 
The canonical divergence satisfies relational expressions called triangular relations for three points $P,Q$ and $R\in M$. This can be regarded as a generalization of the law of cosines in Euclidean space. When a curve $PQ$ and a curve $QR$ are ''orthogonal'', the triangular relation becomes the same expression as the Pythagorean theorem \cite{amari2016information}. This generalized Pythagorean theorem is an important role in the projection theorem.

The present paper aims at studying the divergences in dually flat spaces and their properties and relations. 

First, We introduce a new divergence called ''affine divergence '' which can be expressed as an inner product of $\theta$-coordinate and its dual $\eta$-coordinate.  The affine divergence    
is consistent with the Jeffreys divergence \cite{nielsen2010family,jeffreys1946invariant} for exponential or mixture families and it satisfies triangular relational expression as well as the canonical divergence.

We study the behavior of the affine divergence on $\theta$ and $\eta$-geodesics and show that the affine divergence satisfies the same summation formula as the squared Euclidean distance. We further show that the generalized parallelogram law and the generalized polarization identity hold like Euclidean space for the sum of vectors in $\theta$ and $\eta$-coordinate systems.

Next, we introduce new dual divergences called $\psi$ and $\phi$-divergences by using the sum of vectors of the affine coordinates. These divergences are functions of potentials. The $\psi$-divergence is consistent with the Bhattacharyya distance for exponential families \cite{nielsen2011burbea, bhattacharyya1943measure} and the $\phi$-divergence is consistent with the Jensen-Shannon divergence\cite{lin1991divergence} for mixture families. We derive an inequality that holds between the affine divergence and $\psi,\phi$-divergence and show that it is a generalization of Lin's inequality\cite{lin1991divergence}.
Table1. shows the summary of the properties of divergences.

\begin{table}[h]
\caption{The summary of the properties of divergences}
{\footnotesize In this table, the bold fonts denote the divergences which can be represented by the affine coordinates or potentials. With the canonical divergence as example, the table denotes the canonical divergence is consistent with the KL-divergence for exponential or mixture families and satisfy generalized the law of cosines as the property similar to Euclidean geometry.The blank cells denote unknown.}
\begin{tabular}{l|l|l|l|l}
\cline{2-4}
                                                                                               & \begin{tabular}[c]{@{}l@{}}exponential \\ family\end{tabular}         & \begin{tabular}[c]{@{}l@{}}mixture \\ family\end{tabular}             & \begin{tabular}[c]{@{}l@{}}properties similar\\  to Euclidean geometry\end{tabular}                                                 &  \\ \cline{1-4}
\multicolumn{1}{|l|}{\textbf{\begin{tabular}[c]{@{}l@{}}canonical \\ divergence\end{tabular}}} & \begin{tabular}[c]{@{}l@{}}Kullback-Leibler\\ divergence\end{tabular} & \begin{tabular}[c]{@{}l@{}}Kullback-Leibler\\ divergence\end{tabular} & law of cosines                                                                                                                      &  \\ \cline{1-4}
\multicolumn{1}{|l|}{\textbf{\begin{tabular}[c]{@{}l@{}}affine \\ divergence\end{tabular}}}    & \begin{tabular}[c]{@{}l@{}}Jeffreys\\ divergence\end{tabular}         & \begin{tabular}[c]{@{}l@{}}Jeffreys\\ divergence\end{tabular}         & \begin{tabular}[c]{@{}l@{}}law of cosines\\ summation formula on geodesics\\ parallelogram law\\ polarization identity\end{tabular} &  \\ \cline{1-4}
\multicolumn{1}{|l|}{\textbf{\begin{tabular}[c]{@{}l@{}}psi-\\ divergence\end{tabular}}}       & \begin{tabular}[c]{@{}l@{}}Bhattacharyya\\ distance\end{tabular}      &                                                                       &                                                                                                                                     &  \\ \cline{1-4}
\multicolumn{1}{|l|}{\textbf{\begin{tabular}[c]{@{}l@{}}phi-\\ divergence\end{tabular}}}       &                                                                       & \begin{tabular}[c]{@{}l@{}}Jensen-Shannon\\ divergence\end{tabular}   &                                                                                                                                     &  \\ \cline{1-4}
\end{tabular}
\end{table}

\section{Canonical divergence}
Let $M$ be a dually flat space, we may define dual affine connections $\nabla$, $\nabla^\ast$ and a Riemannian metric $g_{ij}$.
There exist dual affine coordinate systems $\theta$, $\eta$ and dual convex functions $\psi(\theta),\phi(\eta)$ on $M$.
An affine coordinate system $\theta$ corresponds to the connection $\nabla$, and $\eta$ corresponds to the connection $\nabla^\ast$.
$\psi(\theta)$ and $\phi(\eta)$ are in a relationship of Legendre transformation with each other.
\begin{equation}
\label{Legendle}
\phi(\eta)=\theta^i\eta_i-\psi(\theta),
\end{equation}
where we use Einstein notation for $i=(1,2,\cdots,n)$.
$\partial_i$ denotes $\frac{\partial}{\partial\theta^i}$ and $\partial^i$ denotes $\frac{\partial}{\partial\eta_i}$.
Then, equations
\begin{equation}
\label{dif_phi}
\theta^i=\partial^i\phi
\end{equation}
\begin{equation}
\eta_i=\partial_i\psi
\end{equation}
hold.
A relationships between the Riemannian metric and affine coordinates are
\begin{equation}
\label{dif_eta}
g_{ij}=\partial_i\eta_j
\end{equation}
\begin{equation}
\label{dif_theta}
g^{ij}=\partial^i\theta^j.
\end{equation}
For two points $P,Q\in M$, we may define a divergence $D(P\| Q)$ called the canonical divergence as follows \cite{amari2016information,amari2010information}.
\begin{equation}
\label{def_canonical}
D(P\| Q)=\psi(\theta(P))+\phi(\eta(Q))-\theta^i(P)\eta_i(Q)
\end{equation}
The canonical divergence is not symmetric with respect to $P$ and $Q$.
(hereinafter $\psi(P)$ denotes $\psi(\theta(P))$ and $\phi(P)$ denotes $\phi(\eta(P))$.)

For the exponential family $p_\theta(x)=\exp(C(x)+\theta^iF_i(x)-\psi(\theta))$, The Riemannian metric can be expressed as
\begin{eqnarray}
\label{Fisher}
g_{ij}=E[\partial_i l_\theta \partial_j l_\theta]\\
l_\theta=\ln p_\theta(x).
\end{eqnarray}
The right hand side of (\ref{Fisher}) is Fisher information matrix.
The canonical divergence is consistent with the KL-divergence. 
For probability distributions $p(x)$ and $q(x)$, the KL-divergence is 
\begin{equation}
D_{KL}(p\| q)=\sum_ip(x_i)\ln\frac{p(x_i)}{q(x_i)}
\end{equation}
for discrete distribution, and 
\begin{equation}
D_{KL}(p\| q)=\int p(x)\ln\frac{p(x)}{q(x)}dx
\end{equation}
for continuous distribution.
The canonical divergence have distance-like properties(Property 1 and 2), and the properties similar to Euclidean space(Property 3 and 4).\\
\textbf{Property 1.} $D(P\| Q)\geq 0$\\
\textbf{Property 2.} $D(P\| Q)=0 \iff P=Q$\\
\textbf{Property 3.} Triangular relation\\
\begin{equation}
\label{triangular relation}
D(P\| Q)+D(Q\| R)-D(P\| R)=(\eta_i(R)-\eta_i(Q))(\theta^i(P)-\theta^i(Q))
\end{equation}
This formula plays important roles in this paper.
When the dual geodesic connecting $P$ and $Q$ is orthogonal at $Q$ to the dual geodesic connecting $Q$ and $R$, the generalized Pythagorean theorem holds.
\begin{equation}
D(P\| Q)+D(Q\| R)=D(P\| R)
\end{equation}
Because coordinates $\boldsymbol{\theta}\eqdef \{\theta^i\}$ and $\boldsymbol{\eta}\eqdef \{\eta_i\}$ are affine, a curve represented in the form
\begin{equation}
\theta^i(Q(t))=a^it+\theta^i(P)
\end{equation}
is a geodesic for $\nabla$-connection and 
\begin{equation}
\eta_i(Q(t))=a_it+\eta_i(P)
\end{equation}
is a geodesic for $\nabla^\ast$-connection,
where $\{a^i\}$ and $\{a_i\}$ are constant vectors in $\mathbb{R}^n$, $t\in \mathbb{R}$ is a parameter along the geodesic.\\
\textbf{Propertiy 4.} For the point $Q(t)$ on $\theta$-geodesic, $\theta^i(Q(t))=a^it+\theta^i(P)$, the following equation holds.
\begin{equation}
D(P||Q(T))=a^ia^j\int_0^Ttg_{ij}(\theta(Q(t))) dt
\end{equation}

\section{Affine divergence}
\subsection{Definition of the affine divergence\label{Definition of the affine divergence}}

In this section we introduce new divergence called ''affine divergence'' as an inner product of dual affine coordinates.

First, we show the affine divergence satisfies three distance axioms except for the triangle inequality and show the affine divergence is consistent with the Jeffreys divergence for exponential or mixture families.

Second, we show the affine divergence satisfies the properties similar to Euclidean space on geodesics.

Finally, we prove the generalized expansion formula for the sum of vectors, generalized parallelogram law and the generalized polarization identity.

\begin{defi}
\label{semimetric}
A \textbf{semimetric} on $M$ is a function $d:X\times X\rightarrow\mathbb{R}$ that satisfies 
\begin{equation}
\label{semimetric1}
d(P,Q)\geq 0
\end{equation}
\begin{equation}
\label{semimetric2}
d(P,Q)=0 \iff P=Q
\end{equation}
\begin{equation}
\label{semimetric3}
d(P,Q)=d(Q,P).
\end{equation}
\end{defi}

\begin{defi}
We define the \textbf{affine divergence} $D_A:M\times M\rightarrow \mathbb{R}$ as follows.
\begin{equation}
\label{def_geometric}
D_A(P,Q)\eqdef(\eta_i(Q)-\eta_i(P))(\theta^i(Q)-\theta^i(P))
\end{equation}
\end{defi}
\noindent The affine divergence is an inner product of dual affine coordinates.
In a self-dual space($\theta^i=\eta_i$ for all $i$), the affine divergence is consistent with the squared Euclidean distance $\sum_i(\theta^i(Q)-\theta^i(P))^2$.

\begin{prop}
The affine divergence can be expressed as the sum of the canonical divergence.
\begin{equation}
\label{rep_geometric}
D_A(P,Q)=D(P\|Q)+D(Q\|P)
\end{equation}
\end{prop}
\noindent\textbf{Proof.}
The result follows by substituting $P=R$ in the triangular relation(\ref{triangular relation}) and using (\ref{def_geometric}).

\begin{prop}
The affine divergence satisfies semimetric axioms.\\
\noindent\textbf{Proof.}
By using Property 1, Property 2 of the canonical divergence and the definition of the affine divergence(\ref{rep_geometric}), equation (\ref{semimetric1}) and (\ref{semimetric2}) follow. 
Equation (\ref{semimetric3}) is trivial by (\ref{rep_geometric}). 
\end{prop}


For the exponential family $p_\theta(x)=\exp(C(x) +\theta^iF_i(x)-\psi(\theta))$ and the mixture family $p_\eta (x)=p_0(x)+\sum_i \eta_i(p_i(x) - p_0(x))$, $\phi(\eta)=-H(p_\eta)$, the canonical divergence is consistent with the KL-divergence, where $p_i(x)(i=0,1,2,\cdots, n)$ are probability distributions and $H(p)$ is (differential) entropy.
Hence, the affine divergence is consistent with the Jeffreys divergence  $D_J(p,q)\eqdef D_{KL}(p\| q)+D_{KL}(q\| p)$ for exponential or mixture families.
\begin{equation}
D_A(P,Q)=D_{KL}(P\| Q)+D_{KL}(Q\| P)
\end{equation}
The right hand side of this equation is the Jeffreys divergence\cite{nielsen2010family,jeffreys1946invariant}.

\subsection{Properties along geodesics}

We show that the canonical divergence and the affine divergence monotonically increase along $\theta$ and $\eta$-geodesics.
For three points $P,Q,R\in M$ on the same geodesic, we show that the same summation formula as Euclidean space holds.
\begin{prop}
Let $\{a^i\}$ and $\{a_i\}$ be constant vectors in $\mathbb{R}^n$ and $t$ be a parameter $t\in \mathbb{R}$.
When points $P,Q\in M$ are on $\theta$-geodesic 
\begin{equation}
\theta^i(Q(t))=a^it+\theta^i(P),
\end{equation}
the affine divergence can be expressed as
\begin{equation}
D_A(P,Q(T))=Ta^ia^j\int_0^T g_{ij}(\theta(t)) dt.
\end{equation}
When $P,Q\in M$ are on $\eta$-geodesic 
\begin{equation}
\eta_i(Q(t))=a_it+\eta_i(P),
\end{equation}
the affine divergence can be expressed as
\begin{equation}
\label{geometry_int}
D_A(P,Q(T))=Ta_ia_j\int_0^T g^{ij}(\eta(t)) dt.
\end{equation}
Because $g_{ij}$ and $g^{ij}$ are positive definite, the affine divergence monotonically increases with $T$.
\end{prop}
\noindent\textbf{Proof.}
We prove for $\theta$.The same is true of $\eta$.\\
By combining 
\begin{equation}
\frac{d\eta_i(\theta(t))}{dt}=a^j\partial_j \eta_i(\theta(t))
\end{equation}
and (\ref{dif_eta}) yields
\begin{equation}
\label{geodesic_1}
\eta_i(Q(T))-\eta_i(P)=a^j\int_0^T g_{ij}(\theta(t)) dt.
\end{equation}
By substituting (\ref{geodesic_1}) to (\ref{def_geometric}), we have the result.
Equation (\ref{geometry_int}) corresponds to Property 4 of the canonical divergence.

\begin{corollary}
Let $\{a_i\}$ be a constant vector in $\mathbb{R}^n$ and $t$ be a parameter $t\in \mathbb{R}$.
When points $P,Q\in M$ are on $\eta$-geodesic 
\begin{equation}
\eta_i(Q(t))=a_it+\eta_i(P),
\end{equation}
the canonical divergence can be expressed
\begin{equation}
D(P\| Q(T))=a_ia_j\int_0^T dt\int_0^t dt'g^{ij}(\eta(t')). 
\end{equation}
Because $g_{ij}$ is positive definite, the canonical divergence $D(P\| Q(T))$ monotonically increases with $T$.
\end{corollary}
\noindent\textbf{Proof.}
Denoting $\partial_Q^i$ as $\frac{\partial}{\partial \eta_i(Q)}$ and combining the canonical divergence definition (\ref{def_canonical}) and (\ref{dif_phi}), we obtain
\begin{equation}
\partial_Q^i D(P\| Q(t))=\theta(Q(t))^i-\theta(P)^i.
\end{equation}
Hence, we have
\begin{equation}
\frac{dD(P\| Q(t))}{dt}=\frac{1}{t}(\eta(Q(t))_i-\eta(P)_i)(\theta(Q(t))^i-\theta(P)^i)=\frac{1}{t}D_A(P,Q(t)).
\end{equation}
By substituting (\ref{geometry_int}) to this equation and integrating with respect to $t$, we have the result.
\begin{lemma}
\label{division}
Let $t$ be a parameter $t\in\mathbb{R}\setminus \{0,1\}$ and $P,R$ be points on a dually flat space $M$. 
For a point $\boldsymbol{\theta}(Q)=(1-t)\boldsymbol{\theta}(P)+t\boldsymbol{\theta}(R)$ on $\theta$-geodesic, 
\begin{equation}
\label{eq_division}
D(P\| R)=D(P\| Q)+D(Q\| R)+\frac{t}{1-t}D_A(R,Q).
\end{equation}
For a point $\boldsymbol{\eta}(Q)=(1-t)\boldsymbol{\eta}(P)+t\boldsymbol{\eta}(R)$ on $\eta$-geodesic, 
\begin{equation}
D(P\| R)=D(P\| Q)+D(Q\| R)+\frac{1-t}{t}D_A(P,Q).
\end{equation}
\end{lemma}
\noindent\textbf{Proof.}
We prove for $\theta$-geodesic.The same is true of $\eta$-geodesic.\\
By triangular relation (\ref{triangular relation}), we have
\begin{equation}
\label{line}
D(P\| R)=D(P\| Q)+D(Q\| R)+(\eta_i(R)-\eta_i(Q))(\theta^i(Q)-\theta^i(P)).
\end{equation}
By assumption, 
\begin{equation}
\label{eq_ratio}
\theta^i(Q)-\theta^i(P)=\frac{t}{1-t}(\theta^i(R)-\theta^i(Q))
\end{equation}
holds.
By substituting (\ref{eq_ratio}) to (\ref{line}), the result follows.\\
When $t\in(0,1)$, taking into account $D_A(R,Q) \geq 0$, we have
\begin{equation}
D(P\| R)\geq D(P\| Q)+D(Q\| R).
\end{equation}

\begin{corollary}
\label{inv_division}
Let $t$ be  a parameter $t\in\mathbb{R}\setminus \{0,1\}$ and $P,R$ be points on a dually flat space $M$. 
For a point $\boldsymbol{\theta}(Q)=(1-t)\boldsymbol{\theta}(P)+t\boldsymbol{\theta}(R)$ on $\theta$-geodesic,
\begin{equation}
\label{eq_inv_division}
D(R\| P)=D(Q\| P)+D(R\| Q)+\frac{1-t}{t}D_A(Q,P).
\end{equation}
\\
For a point $\boldsymbol{\eta}(Q)=(1-t)\boldsymbol{\eta}(P)+t\boldsymbol{\eta}(R)$ on $\eta$-geodesic,
\begin{equation}
D(R\| P)=D(Q\| P)+D(R\| Q)+\frac{t}{1-t}D_A(R,Q).
\end{equation}
\end{corollary}
\noindent\textbf{Proof.}
We exchange $P$ and $R$ in (\ref{line}), we can prove corollary\ref{inv_division}. in the same way as Lemma\ref{division}.

\begin{theorem}
\label{geometry_division}
Let $t$ be a parameter $t\in\mathbb{R}\setminus \{0,1\}$ and $P,R$ be points on a dually flat space $M$. 
For a point $\boldsymbol{\theta}(Q)=(1-t)\boldsymbol{\theta}(P)+t\boldsymbol{\theta}(R)$ on $\theta$-geodesic or $\boldsymbol{\eta}(Q)=(1-t)\boldsymbol{\eta}(P)+t\boldsymbol{\eta}(R)$ on $\eta$-geodesic,
\begin{equation}
\label{eq_geometry_division}
D_A(P,R)=\frac{1}{t}D_A(P,Q)+\frac{1}{1-t}D_A(Q,R)
\end{equation}
holds.
\end{theorem}
\noindent\textbf{Proof.}
Taking the sum of (\ref{eq_division}) and (\ref{eq_inv_division}), the result follows.\\
Theorem\ref{geometry_division} holds for the points $P,Q,R$ on the same straight line and the squared Euclidean distance.

\subsection{Generalized expansion formula for the sum of vectors, parallelogram law, polarization identity}

For the affine divergence, we show that the generalized law of cosines holds as well as the canonical divergence.
Considering about the sum of vectors $\boldsymbol{\theta}$ or $\boldsymbol{\eta}$, we show that the generalized expansion formula for the sum of vectors, the parallelogram law and the polarization identity hold for the affine divergence.
\begin{defi}
Let $M$ be a dually flat space and $P,Q,R$ be points $P,Q,R\in M$.

We define $\langle , \rangle:M\times M\rightarrow \mathbb{R}$ as
\begin{equation}
\label{def_product}
\langle Q,R \rangle_P\eqdef \frac{1}{2}(\theta^i(Q)-\theta^i(P))(\eta_i(R)-\eta_i(P))+\{Q\leftrightarrow R\}.
\end{equation}
Symbol $\{Q\leftrightarrow R\}$ means replacement of $R$ and $Q$.
\end{defi}
\noindent When $Q=R$, 
\begin{equation}
\langle Q,Q\rangle_P =D_A(P,Q)
\end{equation}
holds. For a self-dual space($\theta^i=\eta_i$ for all $i$), this is consistent with a dot product.

\begin{corollary}(Generalized law of cosines)\\
\label{cor_cosine}
For points $P,Q,R\in M$, 
\begin{equation}
\label{cosine}
D_A(P,Q)+D_A(Q,R)-2\langle P,R \rangle_Q=D_A(P,R)
\end{equation}
holds.
\end{corollary}
\noindent\textbf{Proof.}
By exchanging $P$ and $R$ in triangular relation (\ref{triangular relation}) and taking the sum with original triangular relation (\ref{triangular relation}), we show this corollary.
Equation (\ref{cosine}) is the generalized law of cosines.

\begin{theorem}(Generalized expansion formula for the sum of vectors)\\
\label{vector_sum}
When points $P,Q,R,S\in M$ satisfy $\boldsymbol{\theta}(P)+\boldsymbol{\theta}(R)=\boldsymbol{\theta}(Q)+\boldsymbol{\theta}(S)$ or $\boldsymbol{\eta}(P)+\boldsymbol{\eta}(R)=\boldsymbol{\eta}(Q)+\boldsymbol{\eta}(S)$, the following equations hold.
\begin{equation}
\label{inner_product1}
D_A(P,R)=D_A(P,Q)+D_A(P,S)+2\langle Q,S\rangle _R
\end{equation}
\begin{equation}
\label{inner_product2}
D_A(P,R)=D_A(R,Q)+D_A(R,S)+2\langle Q,S\rangle _P
\end{equation}
\end{theorem}
\noindent\textbf{Proof.}
We prove for $\boldsymbol{\theta}(P)+\boldsymbol{\theta}(R)=\boldsymbol{\theta}(Q)+\boldsymbol{\theta}(S)$.
The same is true of $\boldsymbol{\eta}(P)+\boldsymbol{\eta}(R)=\boldsymbol{\eta}(Q)+\boldsymbol{\eta}(S)$.
By the definition of the affine divergence
\begin{equation}
D_A(P,R)=(\eta_i(R)-\eta_i(P))(\theta^i(R)-\theta^i(P))
\end{equation}
and the assumption
$\boldsymbol{\theta}(P)+\boldsymbol{\theta}(R)=\boldsymbol{\theta}(Q)+\boldsymbol{\theta}(S)$, $D_A(P,R)$ can be expressed as
\begin{equation}
\label{vector_sum1}
D_A(P,R)=(\eta_i(R)-\eta_i(P))(\theta^i(Q)-\theta^i(P)+\theta^i(S)-\theta^i(P)).
\end{equation}
Furthermore, the following relational equation holds.
\begin{eqnarray}
\label{vector_sum2}
(\eta_i(R)-\eta_i(P))(\theta^i(Q)-\theta^i(P))=(\eta_i(R)-\eta_i(Q))(\theta^i(Q)-\theta^i(P))+D_A(P,Q)\nonumber
\\
=(\eta_i(R)-\eta_i(Q))(\theta^i(R)-\theta^i(S))+D_A(P,Q)
\end{eqnarray}
We now use the assumption $\boldsymbol{\theta}(P)+\boldsymbol{\theta}(R)=\boldsymbol{\theta}(Q)+\boldsymbol{\theta}(S)$ again.
In the same way, we obtain the following relational equation.
\begin{eqnarray}
\label{vector_sum3}
(\eta_i(R)-\eta_i(P))(\theta^i(S)-\theta^i(P))=(\eta_i(R)-\eta_i(S))(\theta^i(S)-\theta^i(P))+D_A(P,S)\nonumber
\\
=(\eta_i(R)-\eta_i(S))(\theta^i(R)-\theta^i(Q))+D_A(P,S)
\end{eqnarray}

Substituting (\ref{vector_sum2}) and (\ref{vector_sum3}) to (\ref{vector_sum1}), and using (\ref{def_product}), we prove that (\ref{inner_product1}) holds.
Because the assumption is symmetric with respect to $P$ and $R$, we exchange $P$ and $R$ in (\ref{inner_product1}) and we prove that (\ref{inner_product2}) holds.\\
Theorem \ref{vector_sum} is a generalization of the expansion formula for the squared norm of the sum of  vectors.
\begin{equation}
\|(y-x)+(z-x)\|^2=\|y-x\|^2+\|z-x\|^2+2(y-x,z-x),
\end{equation}
where $( , )$ is a dot product and $\|\cdot\|$ is a Euclidean norm.

\begin{theorem}(Generalized parallelogram law)\\
\label{parallelogram}
When points $P,Q,R,S\in M$ satisfy $\boldsymbol{\theta}(P)+\boldsymbol{\theta}(R)=\boldsymbol{\theta}(Q)+\boldsymbol{\theta}(S)$ or $\boldsymbol{\eta}(P)+\boldsymbol{\eta}(R)=\boldsymbol{\eta}(Q)+\boldsymbol{\eta}(S)$, the following equations holds.
\begin{equation}
D_A(P,Q)+D_A(Q,R)+D_A(R,S)+D_A(S,Q)=D_A(P,R)+D_A(Q,S)
\end{equation}
\end{theorem}
The left hand side is the sum of four sides of rectangle $PQRS$ and the right hand side is the sum of diagonal lines of rectangle $PQRS$.\\
\noindent\textbf{Proof.}
Applying Corollary \ref{cor_cosine} for points $Q, R$ and $S$, we have
\begin{equation}
\label{eq_cosine2}
2\langle Q,S \rangle_R=D_A(R,Q)+D_A(R,S)-D_A(Q,S).
\end{equation}
Substituting (\ref{eq_cosine2}) to (\ref{inner_product1}), we have the result.\\
Theorem \ref{parallelogram} is a generalization of parallelogram law, 
\begin{equation}
2(\|y-x\|^2 + \|z-x\|^2)=\|(y-x)+(z-x)\|^2+\|(z-x)-(y-x)\|^2.
\end{equation}
\begin{corollary}(Generalized polarization identity)\\
\label{polarization}
When points $P,Q,R,S\in M$ satisfy $\boldsymbol{\theta}(P)+\boldsymbol{\theta}(R)=\boldsymbol{\theta}(Q)+\boldsymbol{\theta}(S)$ or $\boldsymbol{\eta}(P)+\boldsymbol{\eta}(R)=\boldsymbol{\eta}(Q)+\boldsymbol{\eta}(S)$, the following equations holds.
\begin{equation}
\label{inner_para}
2(\langle Q,S\rangle _P+\langle Q,S\rangle _R)=D_A(P,R)-D_A(Q,S)
\end{equation}
\end{corollary}
\noindent\textbf{Proof.}\\
By using (\ref{inner_product2}) and (\ref{eq_cosine2}), the result follows.\\
Corollary \ref{polarization} is a generalization of polarization identity,
\begin{equation}
4(y-x,z-x)=\|(y-x)+(z-x)\|^2-\|(z-x)-(y-x)\|^2.
\end{equation}
\begin{corollary}
\label{interior_angle}
When points $P,Q,R,S\in M$ satisfy $\boldsymbol{\theta}(P)+\boldsymbol{\theta}(R)=\boldsymbol{\theta}(Q)+\boldsymbol{\theta}(S)$ or $\boldsymbol{\eta}(P)+\boldsymbol{\eta}(R)=\boldsymbol{\eta}(Q)+\boldsymbol{\eta}(S)$, the following equations holds.
\begin{equation}
\langle Q,S\rangle _P+\langle Q,S\rangle _R+\langle P,R\rangle _Q+\langle P,R\rangle _S=0
\end{equation}
\end{corollary}
\noindent\textbf{Proof.}
Exchanging $(P,R)\leftrightarrow(Q,S)$ in (\ref{inner_para}) and taking the sum with (\ref{inner_para}), we have the result.\\
Corollary \ref{interior_angle} is a generalization that the sum of adjacent interior angles of a parallelogram in Euclidean space is $\pi$.

\section{$\psi$ and $\phi$-divergences}
In this section, we newly introduce dual divergences by using the sum of vectors of affine coordinates. We call these divergence ''$\psi$-divergence'' and ''$\phi$-divergence''.
The $\psi$ and $\phi$-divergences are functions of potential functions.
We show the $\psi$ and $\phi$-divergences are a kind of the skew Jensen divergence\cite{nielsen2010family} and show that the $\phi$-divergence is consistent with the (skew) Jensen-Shannon divergence(JS-divergence) for mixture families. For exponential families, it has been shown that $\psi$-divergence is consistent with the (skew) Bhattacharyya distance \cite{nielsen2011burbea}.
For propbability distributions $p(x)$ and $q(x)$, the JS-divergence $D_{JS}$ and the Bhattacharyya distance $D_B$ are defined as follows.
\begin{equation}
\label{Jensen-Shannon}
D_{JS}(p,q)\eqdef\frac{1}{2}D_{KL}(p\| \frac{p+q}{2}) + \frac{1}{2}D_{KL}(q\| \frac{p+q}{2})
\end{equation}
\begin{equation}
\label{Bhattacharyya}
D_{B}(p,q)\eqdef-\ln(\int dx \sqrt{p(x)q(x)})
\end{equation}
\begin{equation}
\label{Bhattacharyya2}
D_{B}(p,q)\eqdef-\ln\sum_i\sqrt{p_i(x)q_i(x)}, 
\end{equation}
where equation (\ref{Bhattacharyya}) is for continuous cases and equation (\ref{Bhattacharyya2}) is for discrete cases.
Then, we derive an inequality that hold between $\psi$ or $\phi$-divergence and the affine divergence, and show that this inequality is a generalization of Lin's inequality\cite{lin1991divergence} $D_{JS}(p,q)\leq \frac{1}{4}D_J(p,q)$, where $D_J(p,q)\eqdef D_{KL}(p\| q) + D_{KL}(q\| p)$ is the Jeffreys divergence mentioned in section 2.

\begin{prop}
\label{potential_inequality}
Let manifold $M$ be a dual flat space and parameters $a,b$ be $a,b\in\mathbb{R}$.
When points $P,Q,R\in M$ satisfy $\boldsymbol{\theta}(R)=a\boldsymbol{\theta}(P)+b\boldsymbol{\theta}(Q)$ , the following equation holds.
\begin{equation}
\label{eq_potential_inequality_theta}
aD(P\| R)+bD(Q\| R)=(a+b-1)\phi(R)+a\psi(P)+b\psi(Q)-\psi(R)
\end{equation}
When points $P,Q,R\in M$ satisfy $\boldsymbol{\eta}(R)=a\boldsymbol{\eta}(P)+b\boldsymbol{\eta}(Q)$, the following equations holds.
\begin{equation}
\label{eq_potential_inequality_eta}
aD(R\| P)+bD(R\| Q)=(a+b-1)\psi(R)+a\phi(P)+b\phi(Q)-\phi(R)
\end{equation}
\end{prop}
\noindent\textbf{Proof.}\\
We prove for $\boldsymbol{\theta}(R)=a\boldsymbol{\theta}(P)+b\boldsymbol{\theta}(Q)$.The same is true of $\boldsymbol{\eta}(R)=a\boldsymbol{\eta}(P)+b\boldsymbol{\eta}(Q)$.
By (\ref{def_canonical}), we have
\begin{eqnarray}
D(P\| R)=\psi(P)+\phi(R)-\theta^i(P)\eta_i(R)\\ 
D(Q\| R)=\psi(Q)+\phi(R)-\theta^i(Q)\eta_i(R).
\end{eqnarray}
By combining these equations and assumption, we have
\begin{equation}
aD(P\| R)+bD(Q\| R)=a\psi(P)+b\psi(Q)+(a+b)\phi(R)-\theta^i(R)\eta_i(R).
\end{equation}
By using (\ref{Legendle}), the result follows.
If $a\geq 0$ and $b\geq 0$,  (\ref{eq_potential_inequality_theta}) and (\ref{eq_potential_inequality_eta}) are nonnegative.
\begin{defi}
\label{Def_potential_divergence}
Let manifold $M$ be a dual flat space.
For points $P,Q,R\in M$ which satisfy $\boldsymbol{\theta}(R)=(1-\alpha)\boldsymbol{\theta}(P)+\alpha \boldsymbol{\theta}(Q)$
and parameter $\alpha\in (0,1)$,
let us define the $\alpha$-skew $\psi$-divergence as 
\begin{equation}
\label{eq_def_psi_divergence}
D^{(\alpha)}_\psi(P\| Q)\eqdef (1-\alpha) D(P\| R)+\alpha D(Q\| R).
\end{equation}
For points $P,Q,R\in M$ which satisfy $\boldsymbol{\eta}(R)=(1-\alpha)\boldsymbol{\eta}(P)+\alpha\boldsymbol{\eta}(Q)$
and parameter $\alpha\in (0,1)$, let us define the $\alpha$-skew $\phi$-divergence as 
\begin{equation}
D^{(\alpha)}_\phi(P\| Q)\eqdef (1-\alpha)D(R\| P)+\alpha D(R\| Q).
\end{equation}
\end{defi}
From the definition and Property 1 and 2 of the canonical divergence, we can easily confirm the $\alpha$-skew $\psi$ and $\phi$-divergence satisfy
\begin{eqnarray}
D^{(\alpha)}_\psi(P\| Q)\geq 0\\
D^{(\alpha)}_\psi(P\| Q)=0\iff P=Q\\
D^{(\alpha)}_\phi(P\| Q)\geq 0\\
D^{(\alpha)}_\phi(P\| Q)=0\iff P=Q.
\end{eqnarray}

\begin{corollary}
The $\alpha$-skew $\psi$ and $\phi$-divergence can be expressed as follows.
\begin{equation}
\label{eq_psi_divergence}
D^{(\alpha)}_\psi(P\| Q)=(1-\alpha)\psi(P)+\alpha\psi(Q)-\psi(R)
\end{equation}
\begin{equation}
\label{eq_phi_divergence}
D^{(\alpha)}_\phi(P\| Q)=(1-\alpha)\phi(P)+\alpha\phi(Q)-\phi(R)
\end{equation}
\end{corollary}
\noindent\textbf{Proof.}\\
In Proposition \ref{potential_inequality}, substituting $a=1-\alpha$ and $b=\alpha$ and using the definition of the $\alpha$-skew $\psi$ and $\phi$-divergences, the result follows.

Because $\psi(\theta)$ and $\phi(\eta)$ are strictly convex functions, these divergences are a kind of the skew Jensen divergences \cite{nielsen2010family}.
When a dually flat space is self-dual, $\psi$ and $\phi$-divergences are equal to $\frac{\alpha(1-\alpha)}{2}\sum_i (\theta^i(Q)-\theta^i(P))^2$ because potentials $\psi$ and $\phi$ are equal to $\frac{1}{2}\sum_i (\theta^i)^2$ in a self-dual flat space.

\begin{prop}
\label{relation_JS}
Let $p_i(x)(i=0, 1,2,\cdots, n)$ be probability distributions, and let probability distribution $p_\eta (x)$ be the mixture family defined as follows.
\begin{equation}
p_\eta (x)\eqdef p_0(x)+\sum_i \eta_i(p_i(x) - p_0(x))
\end{equation}
Let $D^{(\alpha)}_{JS}$ be the  $\alpha$-skew JS-divergence defined as follows.
\begin{equation}
D^{(\alpha)}_{JS}(p_{\eta(P)}\| p_{\eta(Q)})\eqdef (1-\alpha)D_{KL}(p_{\eta(P)}\| (1-\alpha)p_{\eta(P)}+\alpha p_{\eta(Q)})+\alpha D_{KL}(p_{\eta(Q)}\| (1-\alpha)p_{\eta(P)}+\alpha p_{\eta(Q)})
\end{equation}
For the dually flat space $M$ with the above affine coordinate $\boldsymbol{\eta}$, the $\alpha$-skew $\phi$-divergence is consistent with the  $\alpha$-skew JS-divergence.
\begin{equation}
D^{(\alpha)}_\phi(P\| Q)=D^{(\alpha)}_{JS}(p_{\eta(P)}\| p_{\eta(Q)})
\end{equation}
\end{prop}
\noindent\textbf{Proof.}\\
For mixture families, the potential $\phi(\eta)$ equals to $-H(p_\eta)$, where $H(p)$ is entropy of probability distribution $p$.
Hence, the $\alpha$-skew $\phi$-divergence can be expressed as 
\begin{equation}
\label{entropy_rep}
D^{(\alpha)}_\phi(P\| Q)=H(p_{\eta(R)})-(1-\alpha)H(p_{\eta(P)})-\alpha H(p_{\eta(Q)}).
\end{equation}

Furthermore, for the point $R\in M$ which satisfies $\boldsymbol{\eta}(R)=(1-\alpha)\boldsymbol{\eta}(P)+\alpha\boldsymbol{\eta}(Q)$, the following equation holds.
\begin{align}
p_{\eta(R)}&=p_0(x)+\sum_i \eta_i(R)(p_i(x) - p_0(x))\\ \nonumber
&=(1-\alpha)(p_0(x)+\sum_i \eta_i(P)(p_i(x) - p_0(x)))+\alpha(p_0(x)+\sum_i \eta_i(Q)(p_i(x) - p_0(x)))\\ \nonumber
&=(1-\alpha) p_{\eta(P)}+\alpha p_{\eta(Q)}
\end{align}
Substituting this equation to (\ref{entropy_rep}), we have
\begin{equation}
\label{phi_entropy}
D^{(\alpha)}_\phi(P\| Q)=H((1-\alpha)p_{\eta(P)}+\alpha p_{\eta(Q)}))-(1-\alpha)H(p_{\eta(P)})-\alpha H(p_{\eta(Q)}).
\end{equation}
On the other hand, the $\alpha$-skew JS-divergence for continuous distribution is 
\begin{align}
(1-\alpha)\int p_{\eta(P)}\ln\biggl(\frac{p_{\eta(P)}}{(1-\alpha)p_{\eta(P)}+\alpha p_{\eta(Q)}}\biggr)dx +\alpha\int p_{\eta(Q)}\ln\biggl(\frac{p_{\eta(Q)}}{(1-\alpha)p_{\eta(P)}+\alpha p_{\eta(Q)}}\biggr)dx \\ \nonumber
=H\bigl((1-\alpha)p_{\eta(P)}+\alpha p_{\eta(Q)}))-(1-\alpha)H(p_{\eta(P)}\bigr)-\alpha H(p_{\eta(Q)}).
\end{align}
This equation also holds for discrete distribution. Comparing this equation with (\ref{phi_entropy}), we have the result.

\begin{theorem}
\label{affine_potential_inequality}
Let $P,Q$ be the points in a dually flat space $M$.
For the affine divergence and the $\alpha$-skew $\psi$ or $\phi$-divergence and a parameter $\alpha\in(0,1)$ , the following inequalities hold.
\begin{equation}
\alpha(1-\alpha)D_A(P, Q)\geq D^{(\alpha)}_\psi(P\| Q)
\end{equation}
\begin{equation}
\label{phi_affine_inequality}
\alpha(1-\alpha)D_A(P, Q)\geq D^{(\alpha)}_\phi(P\| Q)
\end{equation}
\end{theorem}
\noindent\textbf{Proof.}\\
Let $R$ be a point which satisfies $\boldsymbol{\theta}(R)=(1-\alpha)\boldsymbol{\theta}(P)+\alpha \boldsymbol{\theta}(Q)$.
From (\ref{eq_def_psi_divergence}) and using $D(P\| Q)\leq D_A(P,Q)$, we have 
\begin{equation}
D^{(\alpha)}_\psi(P\| Q)= (1-\alpha) D(P\| R)+\alpha D(Q\| R)\leq (1-\alpha) D_A(P,R)+\alpha D_A(R,Q).
\end{equation}
Using Theorem \ref{geometry_division}, we have
\begin{align}
 (1-\alpha) D_A(P,R)+\alpha D_A(R,Q)=\alpha(1-\alpha)\biggl(\frac{1}{\alpha}D_A(P,R)+\frac{1}{1-\alpha}D_A(R,Q)\biggr)  \\ \nonumber 
\leq \alpha(1-\alpha)D_A(P,Q)
\end{align}
From this equation, the result follows. The same is true of $\phi$-divergence.

\begin{corollary}(generalized Lin's inequality 1)
Let $p(x)$ be a probability distribution in mixture families.
For the Jeffreys divergence and the $\alpha$-skew JS-divergence and a parameter $\alpha\in(0,1)$, the following inequality holds.
\begin{equation}
\label{JS_inequality}
\alpha(1-\alpha)D_J(p_{\eta(P)}, p_{\eta(Q)})\geq D^{(\alpha)}_{JS}(p_{\eta(P)}\| p_{\eta(Q)}) 
\end{equation}
\end{corollary}
\noindent\textbf{Proof.}\\
By combining Theorem \ref{affine_potential_inequality}, equation $D_A(P,Q)=D_J(p_{\eta(P)}, p_{\eta(Q)})$ as mentioned in the subsection \ref{Definition of the affine divergence} and Proposition \ref{relation_JS}, the result follows.

If $\alpha=\frac{1}{2}$, (\ref{JS_inequality}) is consistent with Lin's inequality.

\begin{corollary} (Generalized Lin's inequality 2)
Let $p(x)$ be a probability distribution in exponential families and $D^{(\alpha)}_{B}(p_{\theta(P)}\| p_{\theta(Q)})$ be the $\alpha$-skew Bhattacharyya distance.
For continuous probability distributions, the $\alpha$-skew Bhattacharyya distance is defined as
\begin{equation}
D^{(\alpha)}_{B}(p\|q)\eqdef -\ln(\int p(x)^{1-\alpha} q(x)^\alpha dx).
\end{equation}
For discrete probability distributions, the $\alpha$-skew Bhattacharyya distance is defined as
\begin{equation}
D^{(\alpha)}_{B}(p\|q)\eqdef -\ln(\sum_i p(x_i)^{1-\alpha} q(x_i)^\alpha).
\end{equation}
For the Jeffreys divergence and the $\alpha$-skew Bhattacharyya distance and a parameter $\alpha\in(0,1)$, the following inequality holds.
\begin{eqnarray}
\label{Bhattacharyya inequality}
\alpha(1-\alpha)D_J(p_{\theta(P)}, p_{\theta(Q)})\geq D^{(\alpha)}_{B}(p_{\theta(P)}\| p_{\theta(Q)}) \\ \nonumber
\end{eqnarray}
\end{corollary}
\noindent\textbf{Proof.}\\
By combining Theorem \ref{affine_potential_inequality} and equations $D_A(P,Q)=D_J(p_{\theta(P)}, p_{\theta(Q)})$ and $D^{(\alpha)}_\psi(P,Q)=D^{(\alpha)}_{B}(p_{\theta(P)}\| p_{\theta(Q)})$, the result follows.

Because the Renyi divergence $D^{(\alpha)}_R(p\|q)$ \cite{renyi1961measures} is defined as $D^{(\alpha)}_R(p\|q)\eqdef \frac{1}{\alpha}D^{(1-\alpha)}_B(p\|q)$, an inequality 
\begin{equation}
\alpha D_J(p_{\theta(P)}, p_{\theta(Q)})\geq D^{(\alpha)}_{R}(p_{\theta(P)}\| p_{\theta(Q)})
\end{equation}
also holds.
These inequalities are generalization of Lin's equality for the Bhattacharyya distance and the Renyi divergence.

\section{Examples}
In this section, we show concrete examples of the affine divergence.
\subsection{Normal distribution}
As representative example of continuous distribution, we think 1-dimentional normal distribution.
\begin{equation}
p_\theta(x)=\frac{1}{\sqrt{2\pi\sigma^2}}\exp(-\frac{(x-\mu)^2}{2\sigma^2})
\end{equation}
The relations between $\theta$, $\eta$ and $\sigma,\mu$ are
\begin{eqnarray}
\theta^1=-\frac{1}{2\sigma^2}\\ \nonumber
\theta^2=\frac{\mu}{\sigma^2}
\end{eqnarray}
\begin{eqnarray}
\eta_1=\sigma^2+\mu^2\\ \nonumber
\eta_2=\mu.
\end{eqnarray}
We calculate the affine divergence of normal distribution as follows.
\begin{equation}
D_A(P,Q)=(\eta_i(Q)-\eta_i(P))(\theta^i(Q)-\theta^i(P))=\frac{({\sigma_Q}^2-{\sigma_P}^2)^2}{2{\sigma_P}^2{\sigma_Q}^2}+(\mu_Q-\mu_P)^2(\frac{1}{2{\sigma_P}^2}+\frac{1}{2{\sigma_Q}^2})
\end{equation}
For points $P,Q,R\in M$ which satisfies $\boldsymbol{\theta}(R)=a\boldsymbol{\theta}(P)+b\boldsymbol{\theta}(Q)$, mean and variance are
\begin{equation}
\label{ave_mu}
\mu_R=\frac{a{\sigma_Q}^2\mu_P+b{\sigma_P}^2\mu_Q}{a{\sigma_Q}^2+b{\sigma_P}^2}.
\end{equation}
\begin{equation}
\label{ave_sigma}
\frac{1}{{\sigma_R}^2}=\frac{a}{{\sigma_P}^2}+\frac{b}{{\sigma_Q}^2}.
\end{equation}
Equation(\ref{ave_sigma}) means weighted harmonic mean and $\mu_R$ is a point to internally divide the straight line $\overline{\mu_P\mu_Q}$ into $b{\sigma_P}^2:a{\sigma_Q}^2$.\\
For the point $P,Q,R\in M$ which satisfies $\boldsymbol{\eta}(R)=a\boldsymbol{\eta}(P)+b\boldsymbol{\eta}(Q)$, an expected value and a variance are
\begin{equation}
{\sigma_R}^2=a{\sigma_P}^2+b{\sigma_Q}^2+a(1-a)\mu_P^2+b(1-b)\mu_Q^2-2ab\mu_P\mu_Q
\end{equation}
\begin{equation}
\mu_R=a\mu_P+b\mu_Q.
\end{equation}

For these quantities, if $a = t, b = 1 - t$ $ (t \in\mathbb{R}) $, Theorem \ref {geometry_division} holds.
In the case of $a=b=1$, Theorem \ref{vector_sum}, the generalized parallelogram law(Theorem \ref {parallelogram}) and the generalized polarization identity(Corollary \ref{polarization}) hold.
\subsection{Binomal distribution}

As representative example of discrete distribution, we think 1-dimentional binomal distribution.
\begin{equation}
p_\theta(k)=\binom nk p^k(1-p)^{(n-k)},
\end{equation}
where $n$ is constant and $p\in[0,1]$.
The relations between $\theta$, $\eta$ and $p$ are
\begin{eqnarray}
\theta=\ln\frac{p}{1-p} \\ \nonumber
\eta=np.
\end{eqnarray}
We calculate the affine divergence of binomal distribution as follows.
\begin{equation}
D_A(P,Q)=(\eta(Q)-\eta(P))(\theta(Q)-\theta(P))=n(p_Q-p_P)\ln\frac{p_Q(1-p_P)}{p_P(1-p_Q)}.
\end{equation}
For the point $P,Q,R\in M$ which satisfies $\boldsymbol{\theta}(R)=a\boldsymbol{\theta}(P)+b\boldsymbol{\theta}(Q)$, the parameter $p$ is
\begin{eqnarray}
p_R=\frac{\exp(\lambda)}{1+\exp(\lambda)}\\ \nonumber
\lambda=\frac{p_P^ap_Q^b}{(1-p_P)^a(1-p_Q)^b}.
\end{eqnarray}
For the point $P,Q,R\in M$ which satisfies $\boldsymbol{\eta}(R)=a\boldsymbol{\eta}(P)+b\boldsymbol{\eta}(Q)$, the parameter $p$ is
\begin{equation}
p_R=ap_P+bp_Q.
\end{equation}
The above operation is defined only for $a,b$ which satisfies $p_R\in[0,1]$. 
For these quantities, if $a = t, b = 1 - t$ $(t \in\mathbb{R}) $, Theorem \ref {geometry_division} holds.
In the case of $a=b=1$, Theorem \ref{vector_sum}, the generalized parallelogram law(Theorem \ref {parallelogram}) and the generalized polarization identity(Corollary\ref{polarization}) hold.
\section{Conclusion}
We have introduced the affine divergence as a function of the affine coordinates and studied the properties similar to Euclidean space in a dually flat space. We have shown the affine divergence satisfies semimetric axioms and triangular relations as well as the canonical divergence.
We also have shown that both the canonical divergence and the affine divergence increase monotonically along geodesics of affine coordinates.
As a property peculiar to the affine divergence on geodesics, we have shown that the same summation formula as the squared Euclidean distance hold for three points on the same geodesic.

Then, for the sum of vectors in affine coordinate systems, we have shown that the generalized expansion formula, the generalized parallelogram law and the generalized polarization identity hold.

Furthermore, we have introduced $\psi$ and $\phi$-divergence which are functions of potentials.
We have derived inequalities between the affine divergence and $\psi$ or $\phi$-divergence and have shown that these inequalities are generalized Lin's inequalities between the Jeffreys divergence and the JS-diveirgence or the Bhattacharyya distance, the Renyi divergence.

It is expected that dually flat spaces further have structures similar to Euclidean space and there are more relations between divergences.
\bibliography{reference_v3}
\end{document}